\documentclass[groupedaddress,superscriptaddress,aps, twocolumn,floatfix]{revtex4-2}

\usepackage{amssymb}
\usepackage{graphicx}
\usepackage{multirow}
\usepackage[table,xcdraw]{xcolor}
\usepackage{soul}
\usepackage{kantlipsum}
\usepackage{amsmath}
\usepackage{amsfonts}
\usepackage{amssymb}
\usepackage{dcolumn} 
\usepackage{braket}
\usepackage{booktabs}
\usepackage{multirow}
\usepackage[table]{xcolor}
\usepackage{dsfont}
\usepackage{times}
\usepackage{epsfig}
\usepackage{xcolor}
\usepackage{tikz}
\usepackage{float}
\usepackage{color}
\usepackage{inputenc}
\usepackage{booktabs}
\usepackage{pgfplotstable}

\usepackage[toc,page]{appendix}
\usepackage{float}

\def\bkappa {\boldsymbol{\kappa}}

\def\btau {\boldsymbol{\tau}}

\def\bupsilon {\boldsymbol{\upsilon}}

\newcommand{\beq}{\begin{equation}}
\newcommand{\eeq}{\end{equation}}
\newcommand{\ba}{\begin{eqnarray}}
\newcommand{\ea}{\end{eqnarray}}
\font\myfont=cmr12 at 12pt

\begin{document}

\preprint{APS/123-QED}




%


\title{Acoustic topological circuitry in square and rectangular phononic crystals}



 \author{Nicolas Laforge}%
 \email{nicolas.laforge@femto-st.fr}
 \affiliation{Institut FEMTO-ST, CNRS UMR 6174, Université de Bourgogne Franche-Comté, 25030 Besançon, France}

 \author{Richard Wiltshaw}
 \affiliation{Department of Mathematics, Imperial College London, London SW7 2AZ, UK}

 \author{Richard V. Craster}%
 \affiliation{Department of Mathematics, Imperial College London, London SW7 2AZ, UK}
 \affiliation{UMI 2004 Abraham de Moivre-CNRS, Imperial College, London SW7 2AZ, UK}
  \affiliation{Department of Mechanical Engineering, Imperial College London, London SW7 2AZ, UK}
  
 \author{Vincent Laude}%
 \affiliation{Institut FEMTO-ST, CNRS UMR 6174, Université de Bourgogne Franche-Comté, 25030 Besançon, France}

 \author{Julio Andrés Iglesias Martínez}%
 \affiliation{Institut FEMTO-ST, CNRS UMR 6174, Université de Bourgogne Franche-Comté, 25030 Besançon, France}

 \author{Guillaume Dupont}%
\affiliation{Aix Marseille Univ, CNRS, Centrale Marseille, IRPHE UMR 7342, 13013 Marseille, France}

 \author{S\'ebastien Guenneau}%
 \affiliation{UMI 2004 Abraham de Moivre-CNRS, Imperial College, London SW7 2AZ, UK}
 
 \author{Muamer Kadic}%
 \affiliation{Institut FEMTO-ST, CNRS UMR 6174, Université de Bourgogne Franche-Comté, 25030 Besançon, France}
 
  \author{Mehul P. Makwana}
 \email{mm107@ic.ac.uk}
 \affiliation{Department of Mathematics, Imperial College London, London SW7 2AZ, UK}
\affiliation{Multiwave Technologies AG, 3 Chemin du Pre Fleuri, 1228 Geneva, Switzerland}

\begin{abstract}
We systematically engineer a series of square and rectangular phononic crystals to create experimental realisations of complex topological phononic circuits. The exotic topological transport observed is wholly reliant upon the underlying structure which must belong to either a square or rectangular lattice system and not to any hexagonal-based structure. The phononic system chosen consists of a periodic array of square steel bars which partitions acoustic waves in water over a broadband range of frequencies ($\sim 0.5$ MHz). An ultrasonic transducer launches an acoustic pulse which propagates along a domain wall, before encountering a nodal point, from which the acoustic signal partitions towards three exit ports. Numerical simulations are performed to clearly illustrate the highly resolved edge states as well as corroborate our experimental findings. To achieve complete control over the flow of energy, power division and redirection devices are required.  The tunability afforded by our designs, in conjunction with the topological robustness of the modes, will result in their assimilation into acoustical devices. 
\end{abstract}
\maketitle
\section{Introduction}

Steering waves around sharp bends, and splitting their energy between different channels, in a robust manner is of interest across many areas of engineering and physics \cite{mekis_high_1996, yariv_coupled-resonator_1999, chutinan_wider_2002, quirrenbach_optical_2001, kok_linear_2007, mitomi_design_1995}. In particular, a great deal of theoretical and experimental work has been devoted to the study of acoustic propagation in defect waveguides (created by removing rows of inclusions) embedded within a two-dimensional phononic crystal \cite{khelifPRB2003b,khelifAPL2004}; introducing defects into an otherwise perfect phononic crystal allows spatially localised acoustic modes to  reside in a band gap. Another mechanism that allows for the manipulation of acoustic energy is the self-collimation effect used in photonics \cite{chigrin_self-guiding_2003,prather_self-collimation_2007} that uses dynamic anisotropic dispersion to steer and partition modes.

\begin{figure}[th!]
	\centering
	\includegraphics[width=8.6cm]{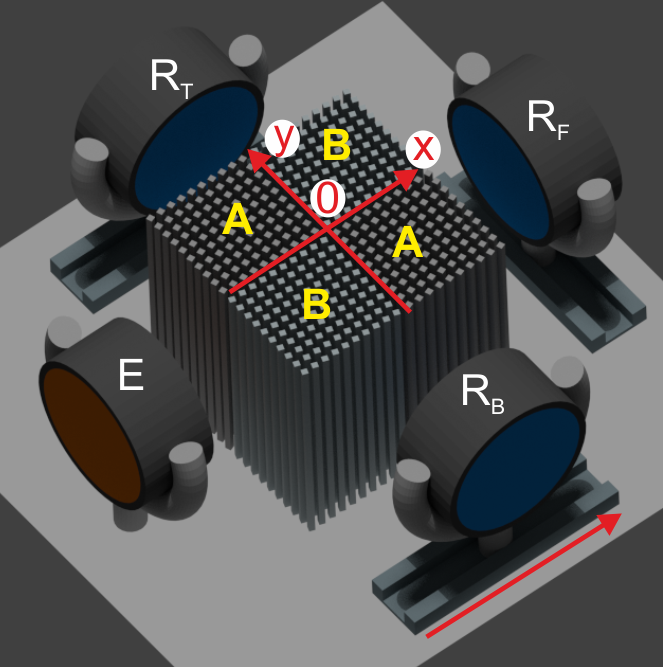} 
	\caption{Experimental setup for the three-way topological beam-splitter: An underwater transducer is used to launch an acoustical pulse, at central frequency $f=0.5$ MHz with a bandwith of $\Delta f=0.25$ MHz, into the four quadrant crystal. A rail-positioned receiver is placed around the crystal to record the transmitter signal in all directions (depicted by the red arrows).}
	\label{fig:setup}
\end{figure}

More recently, and connected to these approaches, there has been intense activity exploring symmetry induced topological-like effects for wave transport that shows little sign of abating \cite{lu_topological_2014, ren_topological_2016, khanikaev_two-dimensional_2017, lu_observation_2016, lu_valley_2016}. These offer an appealing route to passive waveguiding as they do not require convoluted methods for mode coupling between channels, have limited reflection loss and have the added potential of broadband performance amongst other benefits. The physics explored in the present paper centers on edge states which have the same origin as those present in the quantum valley-Hall effect \cite{xiao_valley-contrasting_2007}.  This effect originates from the gapping of Dirac cones via the strategic breaking of spatial symmetries. The regions of quadratic curvature that demarcate the band gap are commonly referred to as the valleys and they are locally endowed with a topological invariant known as the valley Chern number. By attaching two media with opposite valley Chern numbers, broadband chiral edge states are guaranteed to arise; these interface states are commonly known as topological confinement states, kink states, zero modes, or zero-line modes (ZLMs) \cite{lu_valley_2016}. 

Exciting progress is being made in the field of valleytronics \cite{xiao_valley-contrasting_2007, gao_topologically_2017, lu_observation_2016, shalaev_experimental_2017, ma_all-si_2016, makwana_designing_2018, makwana_geometrically_2018, tang_observations_2020, wiltshaw_neumann_2020} and this whole area of research shows much promise. However throughout all of the topological literature there is a gap in experimentally showing how to split energy, over a broadband range of frequencies, in three-directions. Numerous examples of two-way beam splitters have been shown \cite{cha_experimental_2018, he_acoustic_2016, he_two-dimensional_2019, khanikaev_two-dimensional_2017, nanthakumar_inverse_2019, ozawa_topological_2019, qiao_current_2014, schomerus_helical_2010, shen_valley-projected_2019, xia_topologically_2019, yan_-chip_2018, ye_observation_2017, cheng_robust_2016, wu_direct_2017, xia_topological_2017, zhang_manipulation_2018, qiao_electronic_2011}, that leverage the geometry of graphene-like hexagonal structures. The restriction to two-way beam splitting is due to a conservation of a topological charge inherent within graphene-like structures \cite{lu_valley_2016}; to obtain these three-way beam-splitters square, or rectangular, lattice structures must be used and not the hexagonal graphene-like structures that are prevalent in the topological community \cite{cha_experimental_2018, he_acoustic_2016, he_two-dimensional_2019, khanikaev_two-dimensional_2017, nanthakumar_inverse_2019, ozawa_topological_2019, qiao_current_2014, schomerus_helical_2010, shen_valley-projected_2019, xia_topologically_2019, yan_-chip_2018, ye_observation_2017, cheng_robust_2016, wu_direct_2017, xia_topological_2017, zhang_manipulation_2018, qiao_electronic_2011}. Venturing beyond beam-splitters, a longstanding desire within the phononic's sister photonics community has been to mold the flow of light through and around complex opto-electronic “photonic cities”, as shown on the cover of the now classic book by Joannopoulos, Johnson, Winn, and Meade \cite{joannopoulos_photonic_2008}. Herein, we leverage our experimental findings to simulate a phononic circuit that charts the course for future ``phononic cities". Our phononic city comprises of a three-way beam splitter followed by two $\pi/2$ wave steerers (navigation around a $\pi/2$ bend). The resulting pathway is hereafter referred to as a trident-like acoustic circuit.

Due to the underlying topological nature of our designs, wave transport is, both, more robust and also tunable by geometry, hence a wave splitter can be seamlessly changed into a wave steerer by altering a select portion of the structured medium. The geometrical tunability, the topological robustness and the three-way partitioning of energy are the three crucial advantages of our experimental design over competing methods \cite{cha_experimental_2018, he_acoustic_2016, he_two-dimensional_2019, khanikaev_two-dimensional_2017, nanthakumar_inverse_2019, ozawa_topological_2019, qiao_current_2014, schomerus_helical_2010, shen_valley-projected_2019, xia_topologically_2019, yan_-chip_2018, ye_observation_2017, cheng_robust_2016, wu_direct_2017, xia_topological_2017, zhang_manipulation_2018, qiao_electronic_2011}. The phononic circuits created demonstrate the versatility of the underlying mechanisms to efficiently navigate acoustic waves over a broadband range of frequencies. 

In this article, we begin by formulating our problem in Sec. \ref{sec:formulation} before elaborating on the geometrical differences between the square and rectangular phononic systems in Sec. \ref{sec:geometry}. Experimental demonstrations of topological transport along a domain wall, a three-way topological beam-splitter and a trident-like acoustic circuit are shown in Sec. \ref{sec:experiments}; Fig. \ref{fig:setup} illustrates our experimental setup. Finally, in Sec. \ref{sec:conclusion} we draw together some concluding remarks.

\section{Formulation}
\label{sec:formulation}

The phononic crystals we consider are composed of steel rods immersed in water.
Acoustic wave propagation in water is described by the following scalar wave equation,
\begin{equation}\label{eq:1}
\nabla \cdot( \mathbf{\rho}_1^{-1}\nabla p)=
\frac{1}{\kappa}\frac{\partial^2 p}{\partial t^2}
\end{equation}
where $\nabla=(\partial/\partial x_1,\partial/\partial x_2,\partial/\partial x_3)^T$, $^T$ denotes the transpose, $p({\bf x},t)=p(x_1,x_2,x_3,t)$ is the deviation of the acoustic pressure from the ambient pressure, $\mathbf{\rho}_1$ is the homogeneous mass density (in units of kg.m$^{-3}$) of the fluid and $\kappa$ (in units of Pa) is the corresponding homogeneous bulk modulus. 
It is tempting to assume an almost infinite contrast as the bulk modulus of steel is almost two orders of magnitude larger than the bulk modulus of water. However, it has been shown that the elastic waves excited in the solid rods affect the dispersion of the crystal, so that their coupling with acoustic waves in water must be taken into account \cite{laude_phononic_2015}. We model the metal as an elastic body by using Navier's equation:  
\begin{eqnarray}\label{eq:2}
\nabla\cdot \left[ {\mathbb C} :\nabla  \mathbf{u}
 \right]  = \rho_2\frac{\partial^2 \mathbf{u}}{\partial t^2},
\end{eqnarray} where the displacement field ${\bf u}({\bf x},t) = {(u_1({\bf x},t), u_2({\bf x},t), u_3({\bf x},t))}^T$,
and $ {\mathbb C}$ is the rank-4 (symmetric) elasticity tensor with entries $C_{ijkl}=\lambda\delta_{ij}\delta_{kl}+\mu(\delta_{ik}\delta_{jl}+\delta_{il}\delta_{jk})$, $i,j,k,l=1,2,3$, i.e. isotropic,  $\lambda$ and $\mu$ being the Lam\'e parameters (both of them in units of Pa), $\rho_2$ (in units of kg.m$^{-3}$) the mass density of the solid. Equations \eqref{eq:1} and \eqref{eq:2} are related to each other via two coupling conditions at the interface between the fluid and the solid.
First, the pressure in the fluid is continuous with the normal traction in the solid, or $T_{ij} n_j = - p n_i$ where $T = {\mathbb C} :\nabla  \mathbf{u}$ is the stress tensor and ${\bf n}$ is the normal to the interface entering the fluid.
Second, the normal acceleration is continuous and proportional to the gradient of pressure in the fluid according to
\begin{equation}
\mathbf{n}\cdot \left({\rho_1}^{-1} \nabla p\right) = - \mathbf{n}\cdot \frac{\partial^2 \mathbf{u}}{\partial t^2} 
\label{eq:3}
\end{equation}
where the normal ${\bf n}$ now enters the solid.
Therefore, by using this coupled acousto-elastic model we take into account the conversion of pressure waves, in the fluid, into coupled shear and longitudinal elastic waves, in the solid. This allows for us to include subtle physical effects that would otherwise be missed by simpler models \cite{wangPRB2020}; for example, a model that comprises of the scalar wave equation \eqref{eq:1} alongside rigid Neumann boundary conditions.

Hereon in, we assume a time-harmonic dependence $\exp(-\rm{i}\omega t)$, for the pressure field in the fluid phase, and the elastic displacement field, in the solid phase, where $\omega$ is the angular wave frequency in units of radians per second. Moreover, the band spectra, of the doubly-periodic structures under study, are analysed using Floquet-Bloch conditions set on the edges of an elementary cell (edges are chosen within the liquid phase, so that the phase-shift is enforced on the pressure field rather than the displacement field). We do not take into account dissipation (the density and bulk modulus, in Eq. \eqref{eq:1}, and the elasticity tensor and density, in Eq. \eqref{eq:2}, are real valued), therefore we only consider real band spectra ($\omega \in \mathbb{R}$), see e.g. \cite{laude_phononic_2015}. Despite this assumption, the scattering problem, that assumes a Dirac point force in \eqref{eq:1} (the source generating time-harmonic sonic waves is placed within the liquid phase), requires the addition of perfectly matched layers in the fluid phase (with some anisotropic complex valued density and isotropic complex valued  bulk modulus in Eq. \eqref{eq:1}, see \cite{liu_tao_pml_1997}) on either sides of the computational domain to avoid unwanted wave reflections.


\section{Phononic crystal design}
\label{sec:geometry}
\begin{figure}[htb!]
	\centering
	\includegraphics[width=9.25cm]{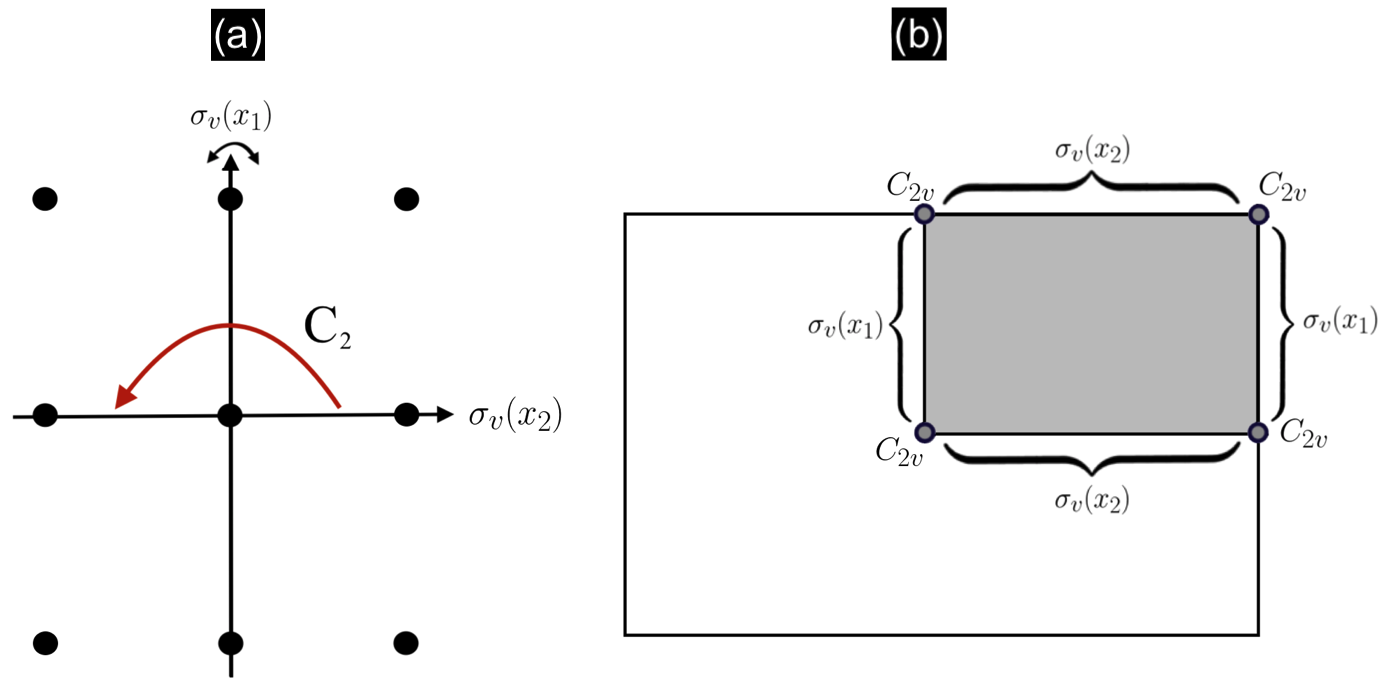} 
	\caption{Panel (a) illustrates the symmetries present within a rectangular structure that belongs to the $pmm$ wallpaper group \cite{dresselhaus_group_2008, Kruthoff_2017}; in our illustration we have assigned $L_1/L_2 < 1$ in Eq. \eqref{eq:generic_basis}. There is a single pair of mirror symmetry lines, denoted by $\sigma_v(x_1)$ and $\sigma_v(x_2)$, as well as a two-fold rotational symmetry. Panel (b) shows the point group symmetries along the high-symmetry lines and high-symmetry points within the rectangular BZ.}
	\label{fig:c2v}
\end{figure}

The cellular structures, that we consider in this article, possess the symmetries shown in Figs. \ref{fig:c4v}(a), \ref{fig:c2v}(a). We chose these structures, over their hexagonal lattice counterparts, due to their inherent ability to localise energy in unique ways \cite{makwana_tunable_2019, proctor_manipulating_2019, APL_Water_Waves, makwana_topological_2019, ungureanu2020localising}; the benefits of this will become more evident in Sec. \ref{sec:experiments}. For the square case, the unrotated cellular structure chosen contains, horizontal, vertical and diagonal mirror symmetries along with four-fold rotational symmetry, Fig. \ref{fig:c4v}(a); whilst for the rectangular case, the unrotated cellular structure solely comprises of a pair of orthogonal mirror symmetries and  a two-fold rotational symmetry, Fig. \ref{fig:c2v}(a). The differences between the dispersive behaviour of waves, propagating through either a square or rectangular lattice, are described in Sec. \ref{sec:dispersion}. Foundational elements of group theory, that we readily use here, can be found in \cite{dresselhaus_group_2008, atkins_molecular_2011}.   
\\

All the space group elements, which leave our chosen square or rectangular lattice invariant, are succinctly written as $\left\{R, \btau \right\}$; where $R$ denotes a point group symmetry element and $\btau$ corresponds to a lattice translation \cite{dresselhaus_group_2008, atkins_molecular_2011}. The square and rectangular periodic structures, under consideration here, belong to a symmorphic space group $G$; hence all $\left\{R, \btau \right\} \in G$ are compound symmetries obtained by combining a point group element and a primitive lattice translation. Specifically, the elements in $G$ are separable, $\left\{R, \btau \right\}=\left\{R, 0 \right\}\left\{\varepsilon, \btau \right\}$; where $\btau$ denotes a Bravais lattice translation. 
\\

If we are at a high-symmetry point in the Brillouin zone (BZ) then $\hat{R}
\bkappa = \bkappa \mod {\bf G}$, for many different $\hat{R}$,  where
${\bf G}$ is an arbitrary reciprocal lattice vector. Each $\hat{R}$ (operator form of $R$),
which satisfies this transformation property at $\bkappa$, belongs to
the \emph{point group} of the wavevector, denoted by $G_{\bkappa}$. The wavevector
group of highest symmetry is $G_{\Gamma}$ (where $\Gamma = (0, 0)$) which is isomorphic to the
factor group $G/T$; $T$ is the translation subgroup. At
high-symmetry points in Fourier space, deterministic degeneracies
may form, which yield a degenerate set of eigenfunctions which correspond to the same
frequency \cite{dresselhaus_group_2008, atkins_molecular_2011, Kruthoff_2017, radha2020topological}.  The vast majority of the valleytronics literature takes advantage of periodic hexagonal lattices, and their symmetry induced Dirac cones at the $KK'$ vertices \cite{Kruthoff_2017, xiao_valley-contrasting_2007}.  Here we deviate away from analysing graphene-like structures and analyse square and rectangular lattice systems which possess non-symmetry repelled Dirac cones; these have to be carefully engineered along specific high-symmetry lines \cite{Kruthoff_2017, makwana_tunable_2019, proctor_manipulating_2019, APL_Water_Waves, makwana_topological_2019}. These structures afford us unique physical properties, such as the partitioning of energy three-ways, the ability to navigate energy around a $\pi/2$ bend whilst preserving the topological protection and the potential to parametrically tune the location of the Dirac cones thereby allowing us to tailor the envelope modulation of the ensuing topological states \cite{makwana_tunable_2019}.

\subsection{Square structure}
\label{sec:square}
\begin{figure}[htb!]
	\centering
	\includegraphics[width=9.25cm]{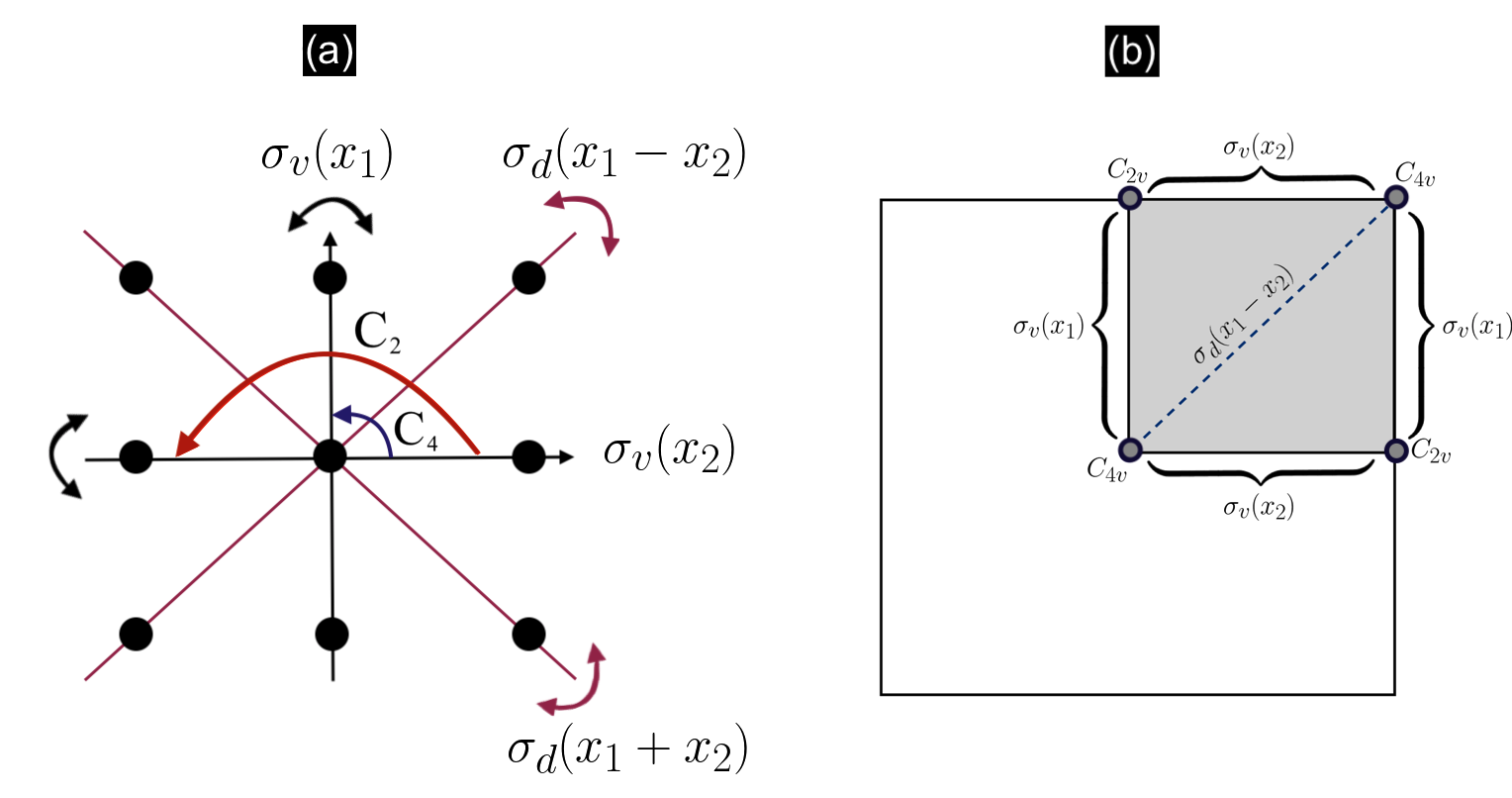} 
	\caption{Panel (a) illustrates the symmetries present within a square structure that belongs to the $p4m$ wallpaper group \cite{dresselhaus_group_2008, Kruthoff_2017}. There are two pairs of mirror symmetry lines, denoted by an orthogonal pair $\sigma_v(x_1), \sigma_v(x_2)$ and another pair of mirror symmetries, $\sigma_d(x_1 + x_2), \sigma_d(x_1 - x_2)$ as well as a set of four-fold rotational symmetries. The dots show the centroid of the cell as well as lattice points that demarcate the cell perimeter. Panel (b) shows the point group symmetries along the high-symmetry lines and at the high-symmetry points, for certain paths contained within the BZ.}
	\label{fig:c4v}
\end{figure}

\begin{figure}[htb!]
	\centering
	\includegraphics[width=7.75cm]{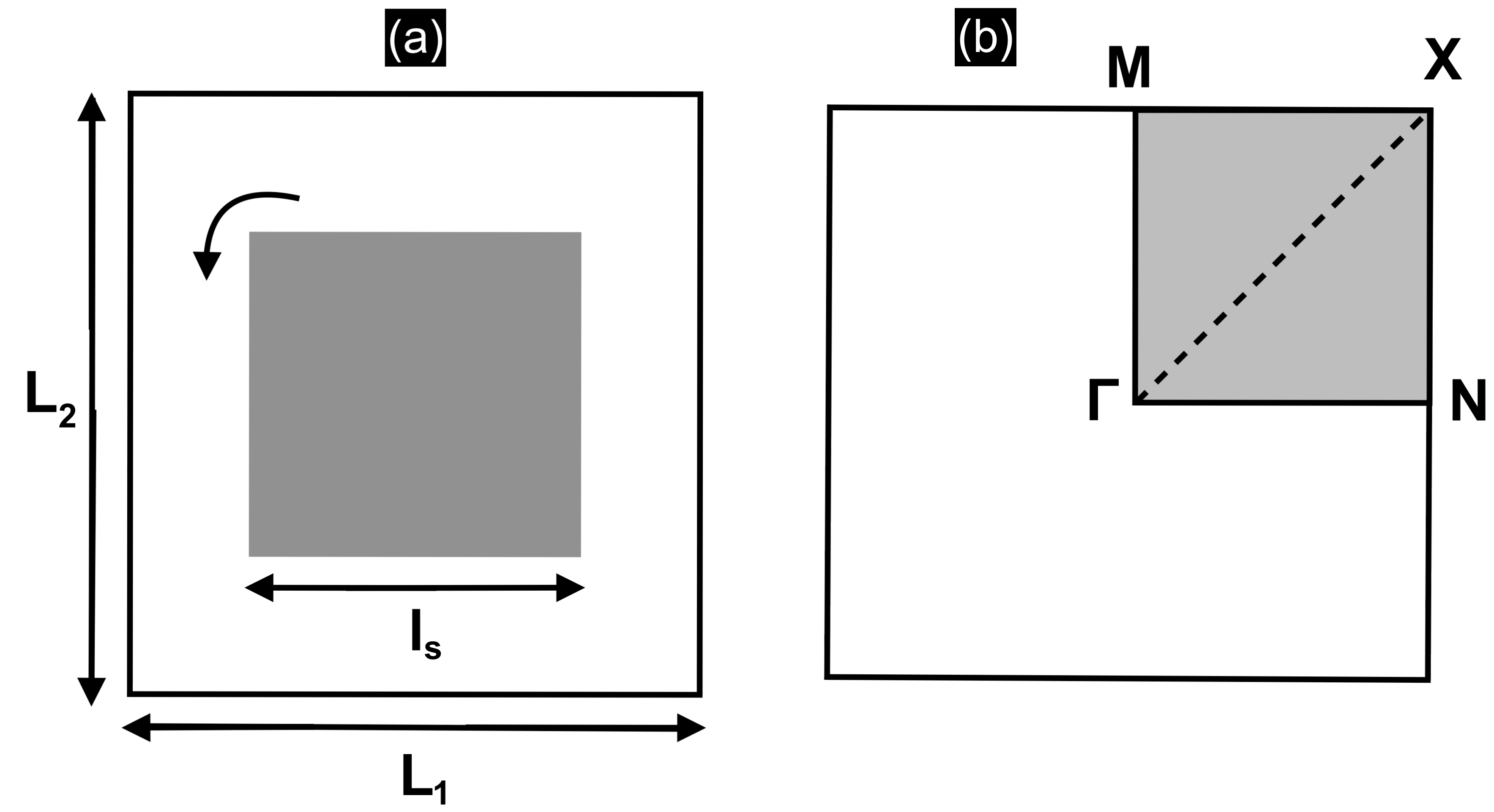} 
	\caption{Panel (a) shows an exemplar physical space cell for an unperturbed square ($L_1 = L_2$) or rectangular ($L_1 \neq L_2$) lattice system; a perturbation is applied when the internal inclusion (grey) is rotated, this breaks all of the reflectional symmetries shown in Figs. \ref{fig:c4v}, \ref{fig:c2v}. We opt to alter the cell dimensions rather than the inclusion dimensions $l_s$ when dealing with a rectangular system. Panel (b) shows the BZ for the square and rectangular cases; $\Gamma N X M$ are the vertices of the square IBZ whilst $\Gamma X M$ are the vertices of the rectangular IBZ.}
	\label{fig:bz_cell}
\end{figure}

The elements within the point group of the square cellular structure, that we consider, are illustrated in Fig. \ref{fig:c4v}(a); they consist of an identity operation ($E$), rotation by $\pm \pi/2$ ($C_4$), rotation by $\pi$ ($C_2$), a pair of orthogonal mirror reflections aligned with the basis vectors ($\sigma_v(x_1), \sigma_v(x_2)$) as well as a pair of diagonal mirror reflections ($\sigma_d(x_1 - x_2), \sigma_d(x_1 + x_2)$). A generic lattice basis vector is given by,
\beq 
\bupsilon = L_1 {\bf i} + L_2 {\bf j}, \quad L_1, L_2 \in \mathbb{R}
\label{eq:generic_basis} \eeq where ${\bf i}, {\bf j}$ are the orthogonal unit vectors and, for a square lattice, $L_1 = L_2$. The physical space cell and the BZ which pertain to Eq. \eqref{eq:generic_basis} are shown in Fig. \ref{fig:bz_cell}. Fig. \ref{fig:c4v}(b) shows the point group symmetries, $G_{\bkappa}$, at several high-symmetry points and high-symmetry lines. Note that the degree of symmetry for a generic $\bkappa$ belonging to, either the high-symmetry lines or high-symmetry points, is greater than for any $\bkappa$ lying off of them. 

\subsection{Rectangular structure}
\label{sec:rectangular}

The rectangular cellular structure's symmetries are a subgroup of the square structure's symmetries. Fig. \ref{fig:c2v}(a) demonstrates that the former's symmetry group merely consists of an identity operation ($E$), rotation by $\pi$ ($C_2$) and a single pair of orthogonal mirror symmetries ($\sigma_v(x_1), \sigma_v(x_2)$). We show in Sec. \ref{sec:experiments} how the absence of the diagonal mirror symmetries, $\sigma_d(x_1 \pm x_2)$, leads to reduced scattering for the rectangular topological modes. A generic rectangular lattice basis vector is given by Eq. \eqref{eq:generic_basis} ($L_1 \neq L_2$). In Sec. \ref{sec:dispersion} we demonstrate how this asymmetry manifests itself in the dispersion diagrams and why this results in reduced topological propagation around a $\pi/2$ bend. Notably, the rectangular structure has lower-order point groups at the high-symmetry points when compared with its square counterpart; there is also no longer a diagonal high-symmetry line, see Fig. \ref{fig:c2v}(b).


\begin{table}[h!]
\begin{tabular}{|l|cc|c|}
\hline
\cellcolor[HTML]{EFEFEF}Classes $\rightarrow$ &                              &                                &                                                                              \\
\cellcolor[HTML]{EFEFEF}IR $\downarrow$       & \multirow{-2}{*}{$E$} & \multirow{-2}{*}{$\sigma$} & \multirow{-2}{*}{\begin{tabular}[c]{@{}c@{}} Parity of basis functions \end{tabular}} \\ \hline
$A$                                         & $+1$                         & $+1$                           & Even                                                                                                                            \\
$B$                                         & $+1$                         & $-1$                           & Odd                                                                                                                                      \\ \hline
\end{tabular}
\caption{$C_{v, d}$ character table}
\label{table:Cs_table}
\end{table}

Character tables, such as Table \ref{table:Cs_table}, are useful for analysing bandstructures along high-symmetry lines \cite{atkins_molecular_2011, dresselhaus_group_2008}. Each row in the table represents a different irreducible representation (IR); these describe the transformational properties of a distinct eigenfunction and are hence associated with separate eigenvalues (distinct bands). Conveniently, the one-dimensional IRs directly tell us about the parity of an eigenmode with respect to a specific symmetry operator. For example, from Table \ref{table:Cs_table}, we see that the action of the mirror symmetry operator ($\hat{\sigma}$) yields either an even ($+$) or odd ($-$) parity eigenmode,
\beq 
\hat{\sigma} P({\bf x}, t) = \pm P({\bf x}, t),
\label{eq:1D_character_ex}\eeq and hence from the signum of the right-hand side we can ascertain which IR an eigenfunction belongs to. Another means to identify, which IR an eigenmode belongs to, is by comparing it against other eigenmodes which lie along the same band but reside at a different $\bkappa$. Due to the continuity of the bands, the IRs belonging to $G_{\bkappa_0}$ will transform adiabatically into the IRs of  $G_{\bkappa_1}$ (where $\bkappa_0$ need not equal $\bkappa_1$). These relationships between different IRs, in different point groups (when $G_{\bkappa_0} \neq G_{\bkappa_1}$), are commonly referred to as compatibility relations \cite{atkins_molecular_2011, dresselhaus_group_2008}.

\subsection{Dispersion diagrams for square and rectangular structures}
\label{sec:dispersion}

\begin{figure}[htb!]
	\centering
	\includegraphics[width=9.05cm]{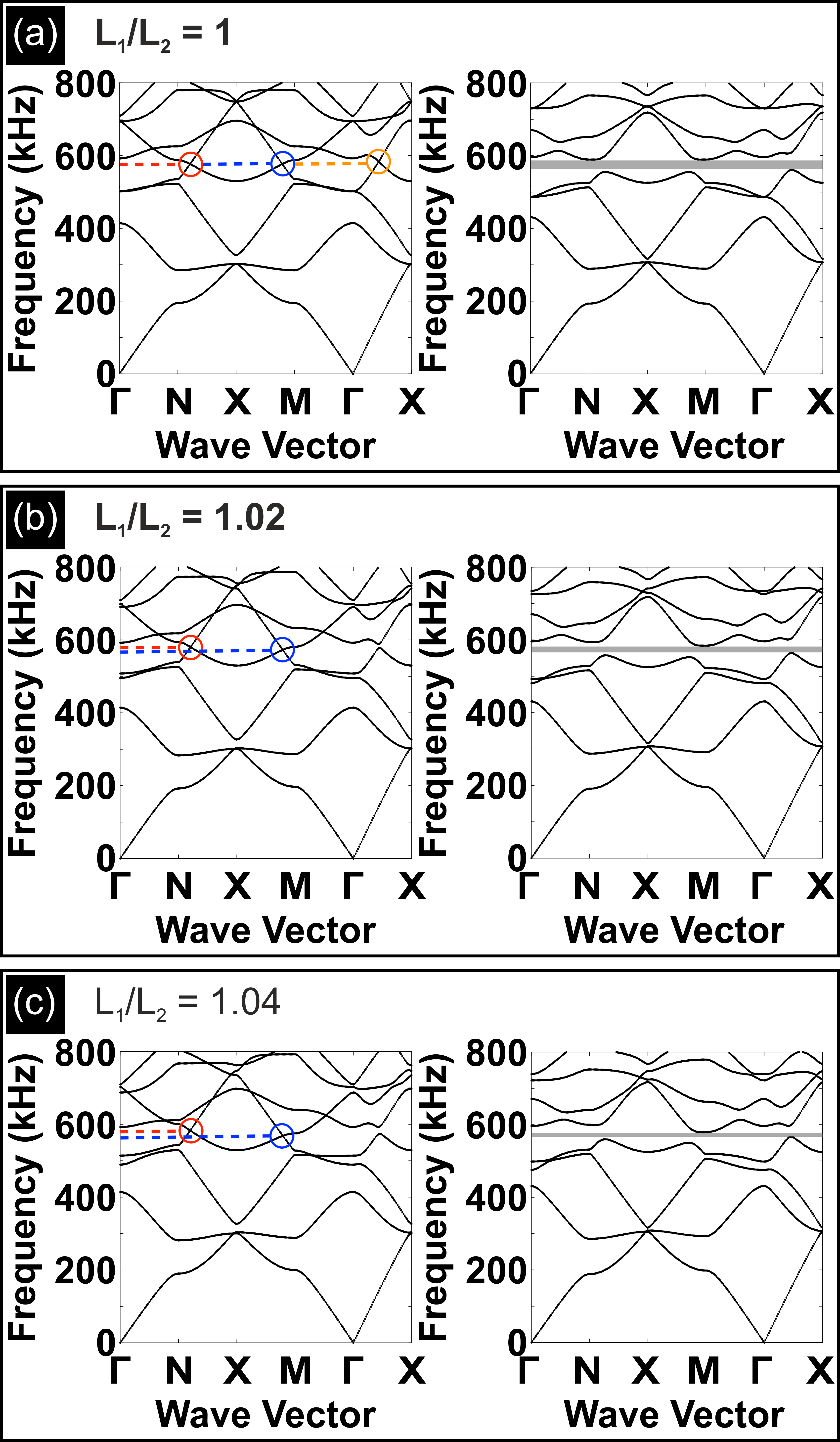} 
	\caption{The dispersion curves for a series of unperturbed (left) and perturbed (right) square and rectangular lattice systems. The asymmetry ratio for each of the panels is (a) $L_1/L_2 = 1$ [$L_1=L_2=3$ mm], (b) $L_1/L_2 = 1.02$ [$L_1=3.03$ mm and $L_2=2.97$ mm], (c) $L_1/L_2 = 1.04$ [$L_1=3.06$ mm and $L_2=2.94$ mm] (see Fig. \ref{fig:bz_cell}). The square inclusion size is $l_s=1.5$ mm. The highlighted regions indicate the locations of the non-symmetry repelled Dirac cones; the red and blue circled Dirac cones belong to the $NX$ and $MX$ perimeter paths of the BZ (Fig. \ref{fig:bz_cell}). Notably, as the asymmetry ratio increases there is a relative frequency shift in their locations. This causes a frequency shift in the ensuing edge states (see Fig. \ref{fig:asymmetry_edge_states}). The orange circled Dirac cone, panel (a), is caused by the $\sigma_d(x_1 \pm x_2)$ mirror symmetry that is only present for the square case (see Fig. \ref{fig:c4v}). The right-sided perturbed dispersion curves are obtained by rotating the inclusion (Fig. \ref{fig:bz_cell}) by $\pi/9$.
	For water, $\kappa=2.25$ GPa and $\rho_1=1000$ kg.m$^{-3}$; for steel $\lambda=115$ GPa, $\mu=77$ GPa, and $\rho_2=7500$ kg.m$^{-3}$.
	}
	\label{fig:dispersion_curves}
\end{figure}

Symmetry generated topological guides leverage the discrete valley degrees of freedom that arise from degenerate extrema in Fourier space. Upon a strategic symmetry reduction, these degeneracies are gapped, leaving a pair of pronounced time-reversal symmetry (TRS) related valleys, that are distinguished by their opposite chiralities \cite{lu_valley_2016}. As we are dealing with a continuous Hermitian system, the bands in the dispersion curves, Fig. \ref{fig:dispersion_curves}, vary continuously, except possibly at accidental degeneracies where mode inversion may occur, which in turn leads to a discontinuity of the intersecting surfaces. Hence, when moving along a continuous band, of simple (real) eigenvalues, the eigenstates continuously transform; departing from the high-symmetry line, the associated IRs, describing the transformation properties of the eigenstates, smoothly transform into the IRs belonging to the eigenstates at a connected high-symmetry point.  

\begin{figure*}[ht!]
	\centering
	\includegraphics[width=17.25cm]{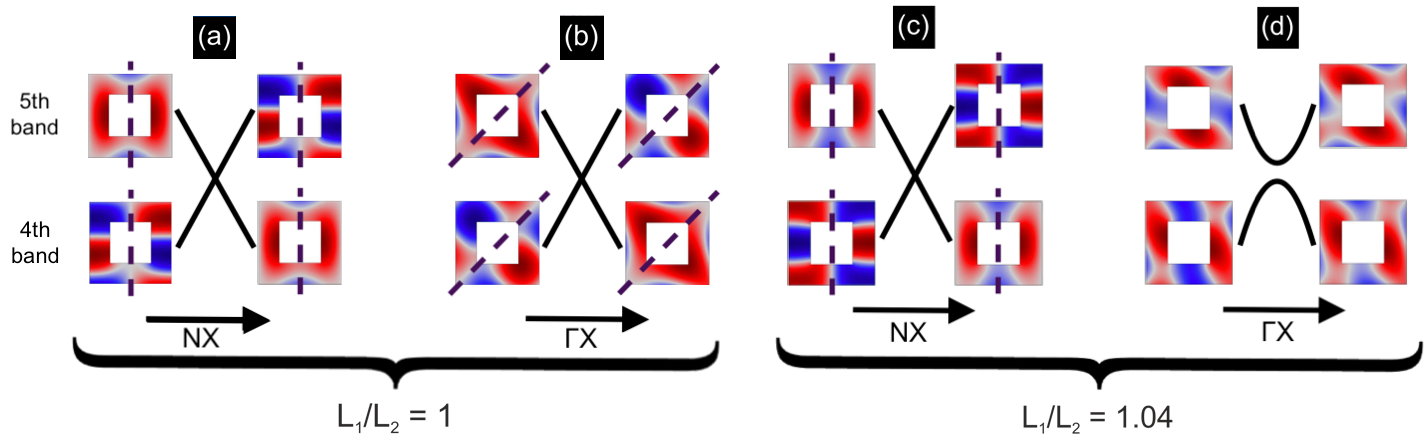} 
	\caption{The eigensolutions, for the $4$th and $5$th bands, of a square ($L_1/L_2 = 1$) and rectangular ($L_1/L_2 = 1.04$) lattice system, at two distinct $\bkappa$'s along each of the paths, $NX$ and $\Gamma X$ in Fig. \ref{fig:dispersion_curves}. The eigensolutions in (a) are taken before and after the $NX$ degeneracy in Fig. \ref{fig:dispersion_curves}(a); notably, we have a pair of opposite parity eigenmodes, with respect to $\sigma_v(x_1)$, and hence they are not symmetry repelled and the bands can cross \cite{makwana_tunable_2019, Kruthoff_2017}. This is also the true, in the square case (b) and the rectangular case (c), where the point group of the path $\Gamma X$ in (b) is $\sigma_d(x_1 - x_2)$. These opposite parity eigenmodes are guaranteed by the point group symmetries, along the $NX$ and $\Gamma X$ paths (Figs. \ref{fig:c4v}(b), \ref{fig:c2v}(b)), and Table \ref{table:Cs_table}. Unlike panels (a), (b) and (c), the eigensolutions in (d) repel due to the hybridisation of even and odd-parity modes; this is due to the absence of $\sigma_d(x_1 \pm x_2)$ symmetry for the rectangular system. 
	}
	\label{fig:eigensolutions}
\end{figure*}

In physical space both the square ($L_1 = L_2$) and the rectangular ($L_1 \neq L_2$) cellular structures, that we consider in Fig. \ref{fig:bz_cell}, have $\sigma_v(x_1)$ and $\sigma_v(x_2)$ symmetries (Figs. \ref{fig:c4v}, \ref{fig:c2v}). This results in the BZ paths $MX$ and $NX$ having $\sigma_v(x_1)$ and $\sigma_v(x_2)$ point group symmetries; therefore the eigensolutions belonging to these paths have a discernible parity (dictated by Table \ref{table:Cs_table}) with respect to these mirror symmetry lines, Figs. \ref{fig:eigensolutions}(a, c). From these eigensolutions we immediately ascertain which bands are repelled from crossing and which are not. Those bands that have an opposite parity, with respect to the mirror symmetry line are able to cross whilst those that have an indeterminate parity do not. Mathematically, this is attributed to the effective Hamiltonian, along $NX$, being diagonalisable and hence the opposite parity bands can be tuned to intersect \cite{ochiai_photonic_2012, makwana_geometrically_2018}.

\cite{makwana_tunable_2019, makwana_topological_2019} demonstrated, using degenerate k$\cdot$p theory, that you could strategically engineer a linear crossing between a pair of $A, B$ bands along the BZ paths, $MX$ and $NX$. System parameters, such as the filling fraction of the inclusion ($l_s$ and/or $L_1, L_2$ in Fig. \ref{fig:bz_cell}), can be tuned such that the opposite parity bands intersect.  This requires two conditions to be satisfied: bands $A, B$ should have opposite slopes and the descending band must be located above the ascending band at some $\bkappa \in NX$ or $MX$. The parametric freedom, afforded by changing the inclusion in geometry or material, allows us to tune the Fourier separation between the TRS-related Dirac cone pairs. The distance between the TRS-related Dirac cones is highly relevant for the transmission properties of the topological guide \cite{makwana_tunable_2019, makwana_wave_2016}. 

Rectangular lattice systems differ from square lattice systems due to the asymmetry between the cell edges ($L_1 \neq L_2$ in Fig. \ref{fig:bz_cell}). This asymmetry manifests itself via a relative shift between the $NX$ and $MX$ Dirac cone locations, see the bandstructures in Fig. \ref{fig:dispersion_curves}. This shift is further enhanced as the ratio $L_1/L_2$ is increased. Another crucial distinction between the square and rectangular structures is the absence of the diagonal mirror symmetry for the latter. The vertical $\sigma_v(x_1)$ and horizontal mirror symmetries, $\sigma_v(x_2)$, yield the non-symmetry repelled Dirac cones along the outer perimeter of the BZ, see Fig. \ref{fig:dispersion_curves}. A similar mechanism occurs, due to the $\sigma_d(x_1 \pm x_2)$ symmetries, whereby the even and odd-parity IRs, belonging to the point group $C_{4v}$ (Fig. \ref{fig:c4v}(b)), transform into the $A, B$ IRs along the path $\Gamma X$ thereby leading to the non-symmetry repelled crossing shown in Figs. \ref{fig:dispersion_curves}(a) and \ref{fig:eigensolutions}(b). Contrastingly, the absence of $\sigma_d(x_1 \pm x_2)$ symmetries for the rectangular structures (Fig. \ref{fig:c2v}) leads to symmetry-induced repulsion along the $\Gamma X$ path in Fig. \ref{fig:dispersion_curves}(b, c); notably, the level repulsion, between the $4$th and $5$th bands, increases as a function of $L_1/L_2$. This phenomenon is because of the indeterminant parity of the eigensolutions in Fig. \ref{fig:eigensolutions}(d). The absence of the $\sigma_d(x_1 - x_2)$ symmetry in Fig. \ref{fig:eigensolutions}(d) when compared to Fig. \ref{fig:eigensolutions}(b) is immediately apparent. This absence results in a pair of non-orthogonal eigensolutions that repel; this repulsion is mathematically attributed to the non-diagonalisability of the effective Hamiltonian for any $\bkappa \in \Gamma X$ \cite{ochiai_photonic_2012, makwana_geometrically_2018, dresselhaus_group_2008}. 


For all the square and rectangular structures, when the inclusion is rotated (Fig. \ref{fig:bz_cell}), all the reflectional symmetries are lost in Fourier space and this breaks open the Dirac points to create the full band-gaps shown in Fig. \ref{fig:dispersion_curves}. Notably, this repulsion of bands is reminiscent of the level repulsion seen in Fig. \ref{fig:eigensolutions}(d). The eigensolutions in Figs. \ref{fig:eigensolutions}(a, b, c) are perturbed into eigensolutions that no longer have a discernible parity and hence repel thereby leaving behind the valleys shown in Fig. \ref{fig:dispersion_curves}. The locally quadratic curves, in the vicinity of the former Dirac cones, carry nonzero Berry curvatures \cite{makwana_tunable_2019, proctor_manipulating_2019, APL_Water_Waves, makwana_topological_2019, ungureanu2020localising} which in turn leads to the generation of valley-Hall edge modes. Those regions on opposite sides of the mirror symmetry line (in Fourier space) carry Berry curvatures with opposite signum \cite{makwana_tunable_2019, proctor_manipulating_2019, APL_Water_Waves, makwana_topological_2019}. The locations of nonzero Berry curvatures, dictate how the geometrically distinct media are stacked \cite{makwana_designing_2018}.


\begin{figure}[htb!]
	\centering
	\includegraphics[width=6.15cm]{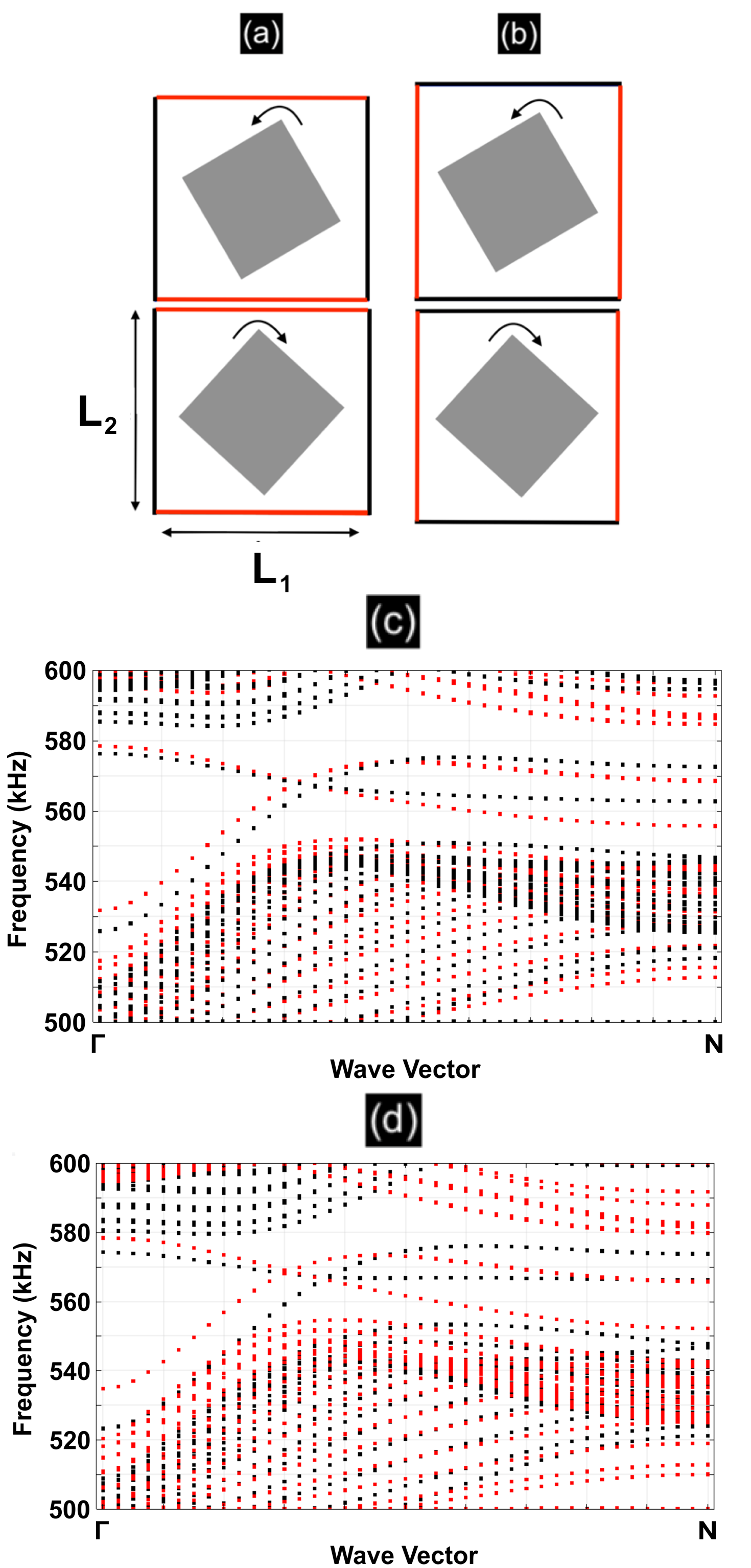} 
	\caption{Panels (a) and (b) show distinct ribbon configurations for a rectangular structure. The rotation of the grey inclusions breaks both the $\sigma_v(x_1), \sigma_v(x_2)$ mirror symmetries shown in Fig. \ref{fig:c2v}(a). Due to $L_1 \neq L_2$ two distinct interfaces (red and black) may be constructed between oppositely perturbed media. Panels (c) and (d), show the edge states, for these interfaces for the rectangular cases $L_1/L_2 = 1.02, 1.04$ respectively; these states reside within the band gaps shown in Figs. \ref{fig:dispersion_curves}(b, c).
	}
\label{fig:asymmetry_edge_states}
\end{figure}


Attaching two topological media, with opposite Berry curvatures yields edge states that are endowed with a designated pseudospin \cite{lu_valley_2016, xiao_valley-contrasting_2007, makwana_designing_2018}. This is achieved by placing one gapped medium, above its reflectional twin, Figs. \ref{fig:asymmetry_edge_states}(a, b); in essence, the stacking in Fourier space results in regions of opposite Berry curvatures overlaying each other, this local disparity ensures the presence of valley-Hall edge modes \cite{qian_theory_2018, wong_gapless_2020, lee-thorp_photonic_2016}. For hexagonal lattice systems, there are 
two distinct orderings of the media which, in turn, creates two distinct interfaces \cite{xiao_valley-contrasting_2007, gao_topologically_2017, lu_observation_2016, shalaev_experimental_2017, ma_all-si_2016, makwana_designing_2018, makwana_geometrically_2018, tang_observations_2020}; one of which supports, only the even modes, and the other, the odd modes. The square and rectangular models differ from this, as their interfaces
support both the even and odd edge modes \cite{makwana_tunable_2019, proctor_manipulating_2019, APL_Water_Waves, makwana_topological_2019, ungureanu2020localising}, Figs. \ref{fig:asymmetry_edge_states}(c, d). This evenness and oddness of the edge modes is inherited from the even and odd bulk modes, Fig. \ref{fig:eigensolutions}. The simplicity of this construction, the apriori knowledge of how to tessellate the two media to produce these broadband edge states, and the added robustness \cite{wong_gapless_2020, orazbayev_chiral_2018} are the main benefits of these topological valley-Hall modes. Specific to our square and rectangular models is the tunable location of the Dirac cones which result in counter-propagating edge states that have a tunable separation in Fourier space; the size of this separation is directly related to the intervalley scattering and hence to the imperviousness of our edge states to shorter-range defects \cite{dong_valley_2017, zhu_design_2018, chen_valley-contrasting_2017}. This tunability is an appealing property as it has been found to be useful for optimising transmission along topological guides as well as for energy-harvesters \cite{makwana_tunable_2019, ungureanu2020localising}. 

The asymmetry between the different cell sides, for the rectangular structure, leads to two distinct interfaces, each of which, hosts even and odd-parity modes (see Fig. \ref{fig:asymmetry_edge_states}). This differs from the square structure, that possesses only a single unique interface, upon which both even and odd-parity modes reside \cite{makwana_tunable_2019, proctor_manipulating_2019, APL_Water_Waves, makwana_topological_2019, ungureanu2020localising}. The asymmetric edges in the rectangular case results in two sets of edge states that do not have a complete frequency overlap (Fig. \ref{fig:asymmetry_edge_states}). This physically affects the modal coupling between the incoming and outgoing leads when dealing with a $\pi/2$ wave steerer \cite{makwana_tunable_2019, mekis_90bend_1996} as the frequency range of operation will be limited when compared to the square case. Despite this, the square case has a significant drawback in that the presence of the diagonal mirror symmetry can lead to enhanced intervalley scattering when compared to the rectangular case (see Sec. \ref{sec:experiments}).


\section{Experimental validation of phononic crystal designs}
\label{sec:experiments}


The system we experimentally study is shown in Fig. \ref{fig:setup} and consists of a square (or rectangular) array of $20 \times 20$ square steel rods fully submerged in water.
The size of the rods is $l_s = 1.5$ mm and the lattice constant is $\approx 3$ mm. 
The rods are aligned and held in place with two 3D-printed parallel perforated plates.
The square array is transformed into a rectangular array by increasing the distance between the rods in the $x_1$-direction, this is equivalent to allowing $L_1$ to change whilst keeping $L_2$ fixed in Fig. \ref{fig:bz_cell}.
In practice, new alignment plates are printed each time the spacing needs to be changed. A  Gaussian-shaped pressure field is emitted by a one inch ultrasonic underwater transducer. 
The central frequency is 600 kHz and the full width at half maximum exceeds 500 kHz.
Therefore, the measurement bandwidth exceeds the band gap width and the pulse-echo technique yields the spectral transmission from the Fourier transform of a single measurement.
The length of the rods is $10$ cm and is thus significantly larger than the transducer.
The pressure field is spatially resolved and measured at different positions outside the array. The outgoing signal is spatially convoluted due to the finite size of the receiver. This signal is then deconvoluted to allow for a direct comparison against the numerical results. 

\subsection{Acoustic topological transport along a domain wall}
\label{sec:domain_wall}
\begin{figure}[htb!]
	\centering
	\includegraphics[width=8.5cm]{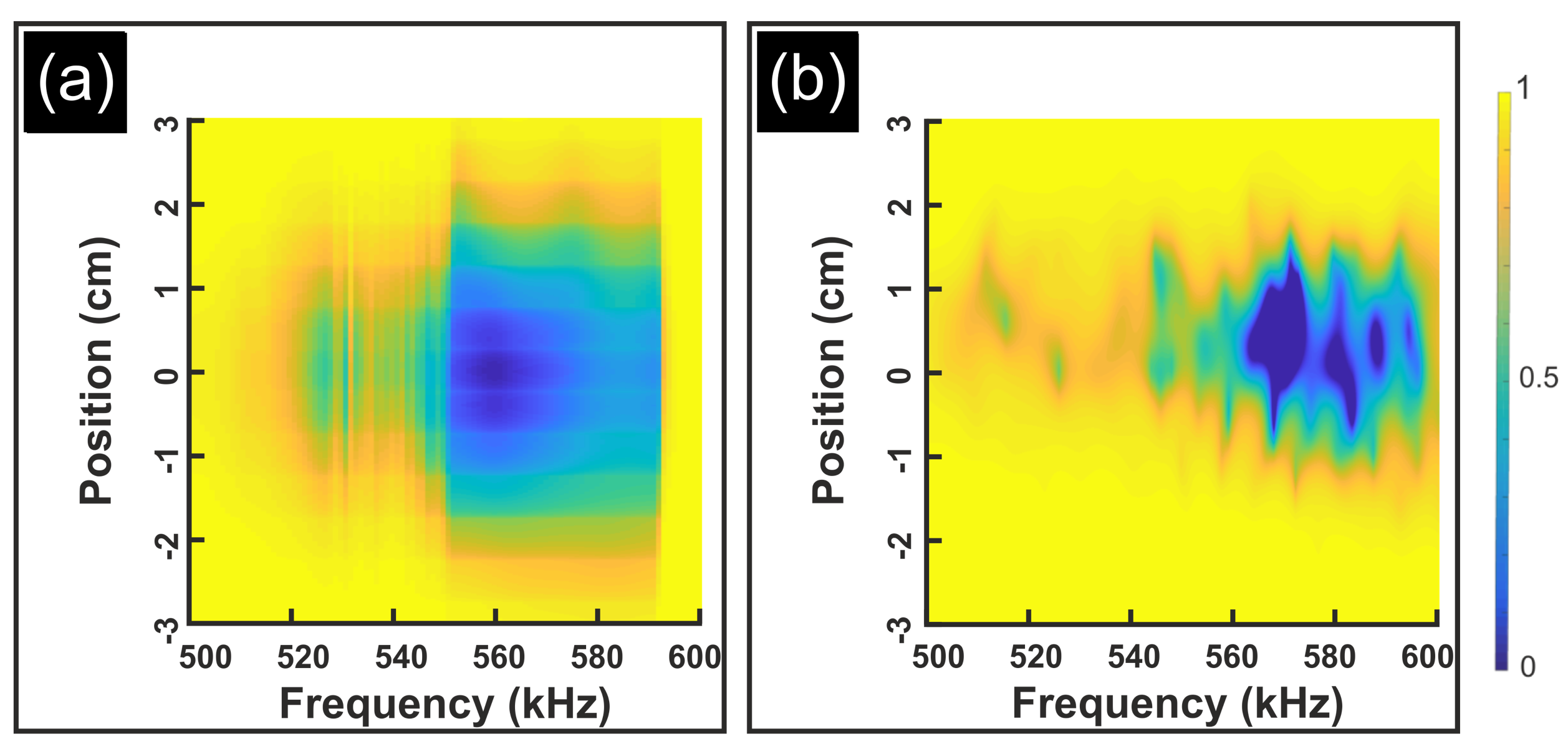} 
	\caption{Numerical (a) and experimental (b) validation for the band gap frequency range associated with the square structure dispersion curves in Fig. \ref{fig:dispersion_curves}(a)}
	\label{fig:band_gap}
\end{figure}

From Fig. \ref{fig:dispersion_curves}, we saw that when the orientation angle of the square inclusion is set to zero ($\theta$ = 0), the point groups along a series of high-symmetry lines are $\sigma_v(x_1), \sigma_v(x_2)$ or $\sigma_d(x_1 - x_2)$; using these mirror symmetries, Dirac points are engineered to exist along the paths $NX, MX$ for square and rectangular structures and also, along $\Gamma X$,  for square structures (see Fig. \ref{fig:dispersion_curves}). Breaking the mirror symmetries, as we do by rotating by $\theta= \pm \pi/9$, gaps the Dirac point to open up a band gap; for the
square structure the complete band gap ranges between {\color{black} $550$ kHz} to {\color{black} $590$ kHz} and this is numerically verified in Fig. \ref{fig:band_gap}(a).
Experimentally, the complete band gap ranges between {\color{black} $555$ kHz} to {\color{black} $595$ kHz}, see Fig. \ref{fig:band_gap}(b).

\begin{figure}[htb!]
	\centering
	\includegraphics[width=8.15cm]{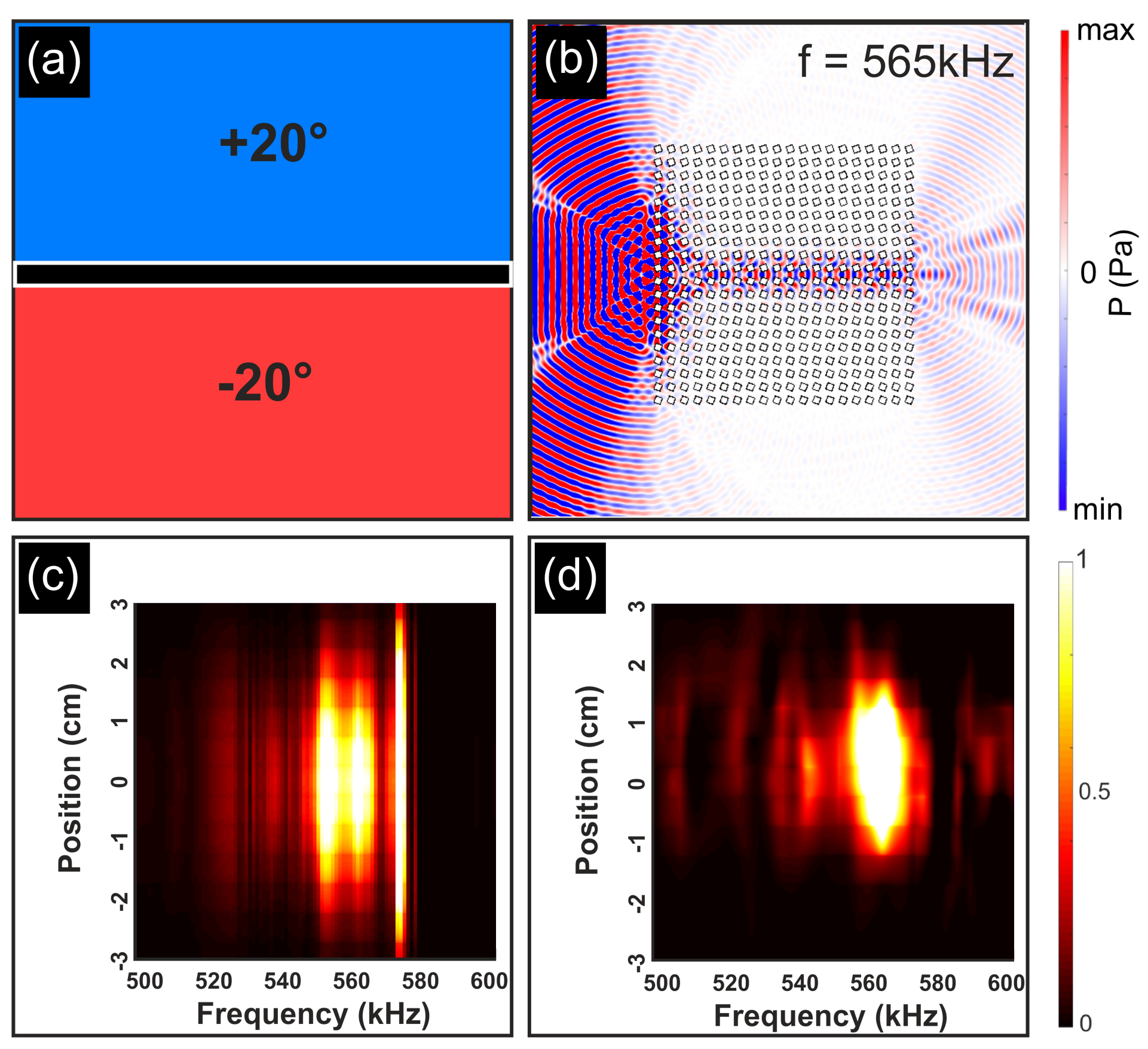} 
	\caption{Simulations and experiments of valley-Hall edge states for a square array. Oppositely perturbed media are placed above and below each other (a), an exemplar even-parity ZLM, for $f = 565$kHz, is shown in (b). Numerical computation of the ZLM frequency range (c) shows good agreement with the experimental range (d). These ranges lie within the bulk band gaps in Fig. \ref{fig:band_gap}
	}
	\label{fig:zlm}
\end{figure}

Numerically we consider an arrangement, similar to that shown in Figs. \ref{fig:asymmetry_edge_states}(a, b) albeit now for the square structure, therefore $L_1 = L_2$ and the ribbon configuration in Fig. \ref{fig:asymmetry_edge_states}(a) matches that in Fig. \ref{fig:asymmetry_edge_states}(b). This results in a single pair of edge states that lie along a single unique interface \cite{makwana_tunable_2019, proctor_manipulating_2019, APL_Water_Waves, makwana_topological_2019}. Both broadband ZLMs exist over a simultaneous frequency range and are distinguished by their opposite parity modal patterns \cite{makwana_tunable_2019, proctor_manipulating_2019, APL_Water_Waves, makwana_topological_2019, ungureanu2020localising}.  This result is predictable via group theoretical considerations; consider a ribbon configuration, similar to that in Figs. \ref{fig:asymmetry_edge_states}(a, b), that is infinite in extent, in the $x_1$ and $x_2$ directions. This structure belongs to the $p1m1$ frieze group \cite{dresselhaus_group_2008} and hence Table \ref{table:Cs_table} guarantees a pair of even and odd-parity modes. The parities are taken, with respect to the mirror symmetry line, which in this instance is the interface. The numerical and experimental demonstration of topological transport along a domain wall, for our particular setup, is shown in Fig. \ref{fig:zlm}. Notably for, both, the bulk band gap (Fig. \ref{fig:band_gap}) and ZLM results (Fig. \ref{fig:zlm}), the configurations were constructed using the same experimental equipment. This is convenient, as we can essentially re-use the same equipment to investigate even more complex topological domains; this allows us to navigate acoustic energy in vastly different directions and with different underlying properties with ease. 

\begin{figure}[h!]
	\centering
	\includegraphics[width=8.75cm]{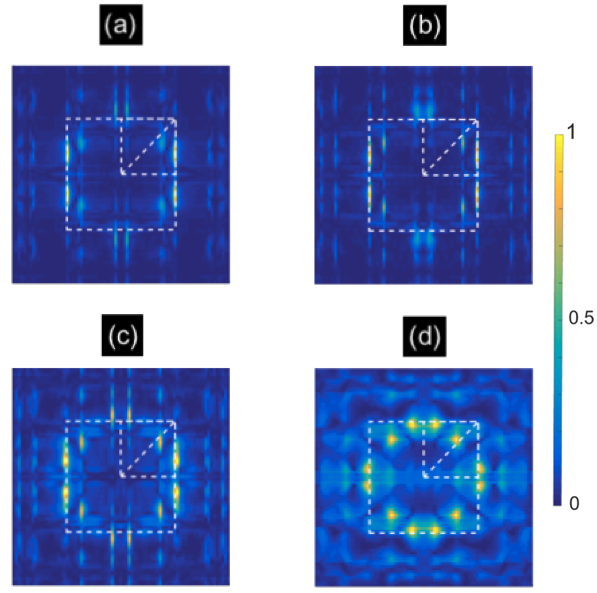} 
	\caption{FFTs of several ZLM modes for the rectangular ($L_1/ L_2 = 1.02$) structure are shown. Panels (a, b, c, d) correspond to frequencies {\color{black} $f = 555, 560, 565, 575$ kHz}. Crucially, as you converge to frequencies near the band gap edge in Fig. \ref{fig:dispersion_curves}(b) the modal excitation along the diagonal mirror symmetry lines is enhanced.
	}
	\label{fig:FFT}
\end{figure}
 
The Fast Fourier Transforms (FFTs) of the wavefield, shown in Fig. \ref{fig:FFT}, exemplify a crucial difference between the square and rectangular structures. The FFTs here correspond to ZLM propagation between two geometrically distinct rectangular arrays (Fig. \ref{fig:asymmetry_edge_states}) within the {\color{black} frequency range $555:575$ kHz}. This range incorporates the bulk band gap shown in Fig. \ref{fig:dispersion_curves}(b) as well as states that lie close to the propagating regime. Importantly, as the frequency is increased the dominant modal excitations are no longer localised to the valleys that lie along the perimeter of the BZ. In Figs. \ref{fig:FFT}(c, d) the diagonal modes along $\sigma_d(x_1 \pm x_2)$ become excited. By using a rectangular array, where there is no diagonal mirror symmetry as opposed to a square array, we are able to obtain a clear set of frequencies for which this diagonal mode excitation is absent. This is beneficial as the topological protection of valley-Hall states arises from, both, the opposite chirality of opposite propagating modes and the intervalley Fourier separation between the counter-propagating states \cite{dong_valley_2017, zhu_design_2018, chen_valley-contrasting_2017}. If there is additional excitation, for example along the $\sigma_d(x_1 \pm x_2)$ lines, this will inevitably lead to enhanced scattering and hence weaker protection.

\subsection{Three-way beam splitting for topological acoustic modes}
\label{sec:three_way}
Over 25 years ago,  high transmission was achieved in 2D photonic crystals using line defects, produced by the removal of circular inclusions, within an otherwise perfectly periodic crystal \cite{mekis_90bend_1996}. Propagation, around a $\pi/2$ bend, with high transmission was achieved. The simulations of the bent waveguide, in \cite{mekis_90bend_1996}, demonstrated $>90$ percent transmission, which is significantly higher than the $30$ percent transmission achieved in bent dielectric waveguides.
Drawbacks of the approach in \cite{mekis_90bend_1996} are that the waveguide is highly sensitive to crystal imperfections inside the defect lines, as well as to small changes in the operating frequency. Comparatively, topological waveguides \cite{lu_topological_2014} allow for highly efficient robust wave transmission, that is more impervious to crystal imperfections, and is valid over a broadband range of frequencies.

Throughout all of the topological literature numerous examples of two-way beam splitters have been shown \cite{cha_experimental_2018, he_acoustic_2016, he_two-dimensional_2019, khanikaev_two-dimensional_2017, nanthakumar_inverse_2019, ozawa_topological_2019, qiao_current_2014, schomerus_helical_2010, shen_valley-projected_2019, xia_topologically_2019, yan_-chip_2018, ye_observation_2017, cheng_robust_2016, wu_direct_2017, xia_topological_2017, zhang_manipulation_2018, qiao_electronic_2011}, that leverage the geometry of graphene-like hexagonal structures. The restriction to two-way energy-splitting is due to a conservation of a topological charge near the $KK'$ valleys and is inherent within graphene-like structures; to obtain these three-way energy-splitters square, or rectangular, structures must be used and not the graphene-like structures that are readily found in the topological community \cite{xiao_valley-contrasting_2007, gao_topologically_2017, lu_observation_2016, shalaev_experimental_2017, ma_all-si_2016, makwana_designing_2018, makwana_geometrically_2018, tang_observations_2020, wiltshaw_neumann_2020}. 

\begin{figure*}[ht!]
	\centering
	\includegraphics[width=18.15cm]{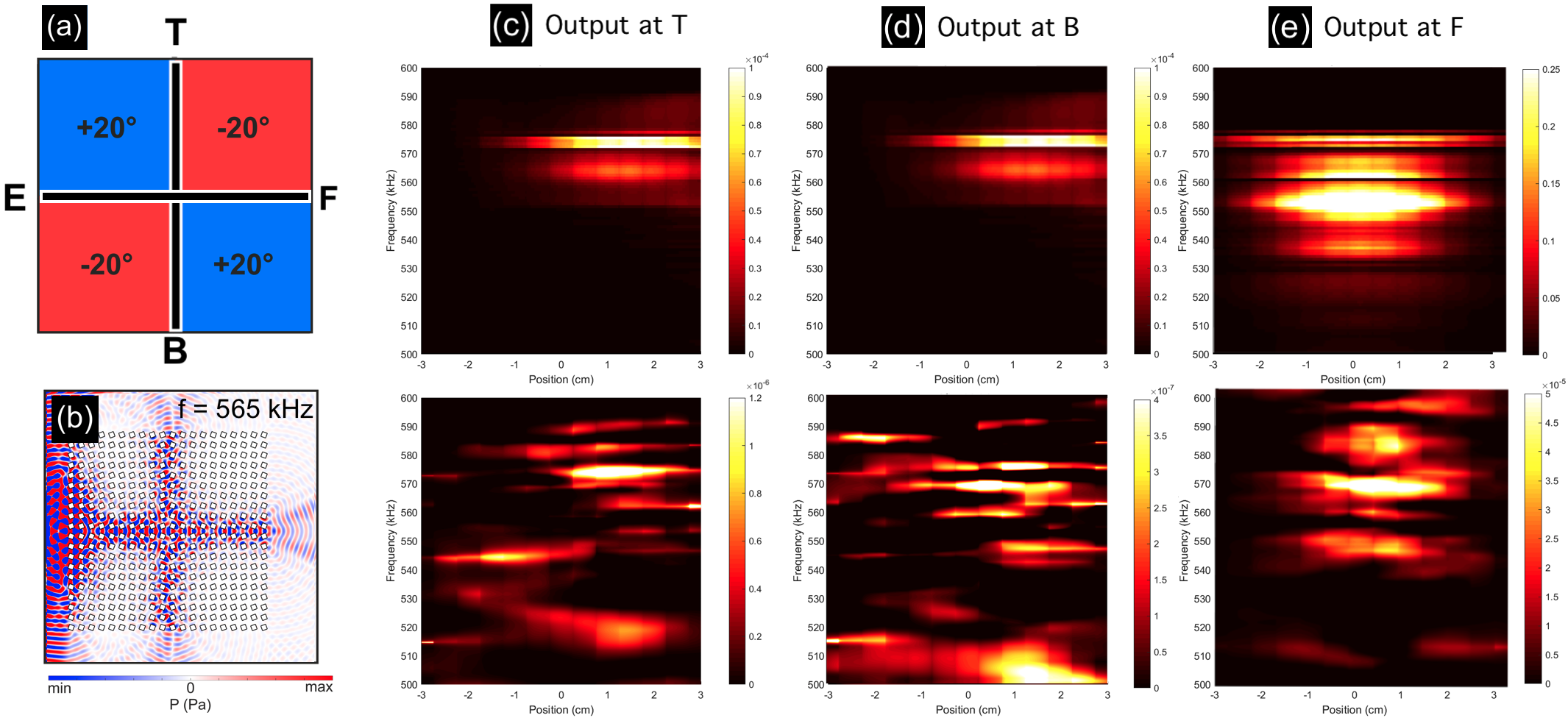} 
	\caption{Three-way splitter for square lattice: (a) shows the four quadrant configuration of our topological domain which resembles Fig. \ref{fig:setup}, the emitter (E) and the top (T), bottom (B) and forward (F) output ports are indicated; the perturbation angular rotation of the square rods alternates between $+\pi/9$ and $-\pi/9$ radians as shown. An exemplar scattering simulation, clearly illustrating the effect, is shown in (b) for frequency $565$ kHz. The numerical simulations (top row) and the experimental measurements (bottom row) for the outgoing ports are shown in (c, d, e). From the simulations we clearly see that there exists a simultaneous frequency range $550:580$ kHz in which all three outgoing ZLMs are ignited. 	
	}
	\label{fig:square_three_way}
\end{figure*}

The configuration for the three-way beam splitter, for a square lattice, is shown in Fig. \ref{fig:square_three_way}(a).  A Gaussian pulse, incident from the left-sided emitter (E), couples with the leftmost interface before being partitioned three-ways towards the T, B and F output ports. A numerically computed scattering solution, for a Gaussian source, is shown in Fig. \ref{fig:square_three_way}(b); the wave-splitter can be seamlessly changed into a wave-steerer (navigation around a $\pi/2$ bend) by altering the lower-right quadrant of the structured medium (from a $``+"$ angular perturbation to a $``-"$).


 For a mode to couple, from one ZLM to another, their pseudospins and their $\bkappa$'s must match \cite{makwana_designing_2018}. For either, the square or rectangular case, these conditions are satisfied due to the relationship between interfaces that separate oppositely perturbed media. For example, Fig. \ref{fig:asymmetry_edge_states}(a) or (b), shows an interface constructed from a positively perturbed medium lying above a negatively perturbed medium; a rightward propagating wave along one of these interfaces would be equivalent to a leftward propagating wave along an interface that separates a negatively perturbed medium above a positively perturbed medium. It is precisely this relationship that allows for the square and rectangular models to partition energy three-ways whilst the hexagonal models cannot . For the latter, the two stackings of the media results in a distinct pair of edge state curves, hence an incident mode along one cannot easily couple to an outgoing ZLM along another, as there is a pseudospin or $\bkappa$ mismatch (see summary table in  \cite{makwana_tunable_2019}). 
 
 \begin{figure*}[ht!]
	\centering
	\includegraphics[width=18.15cm]{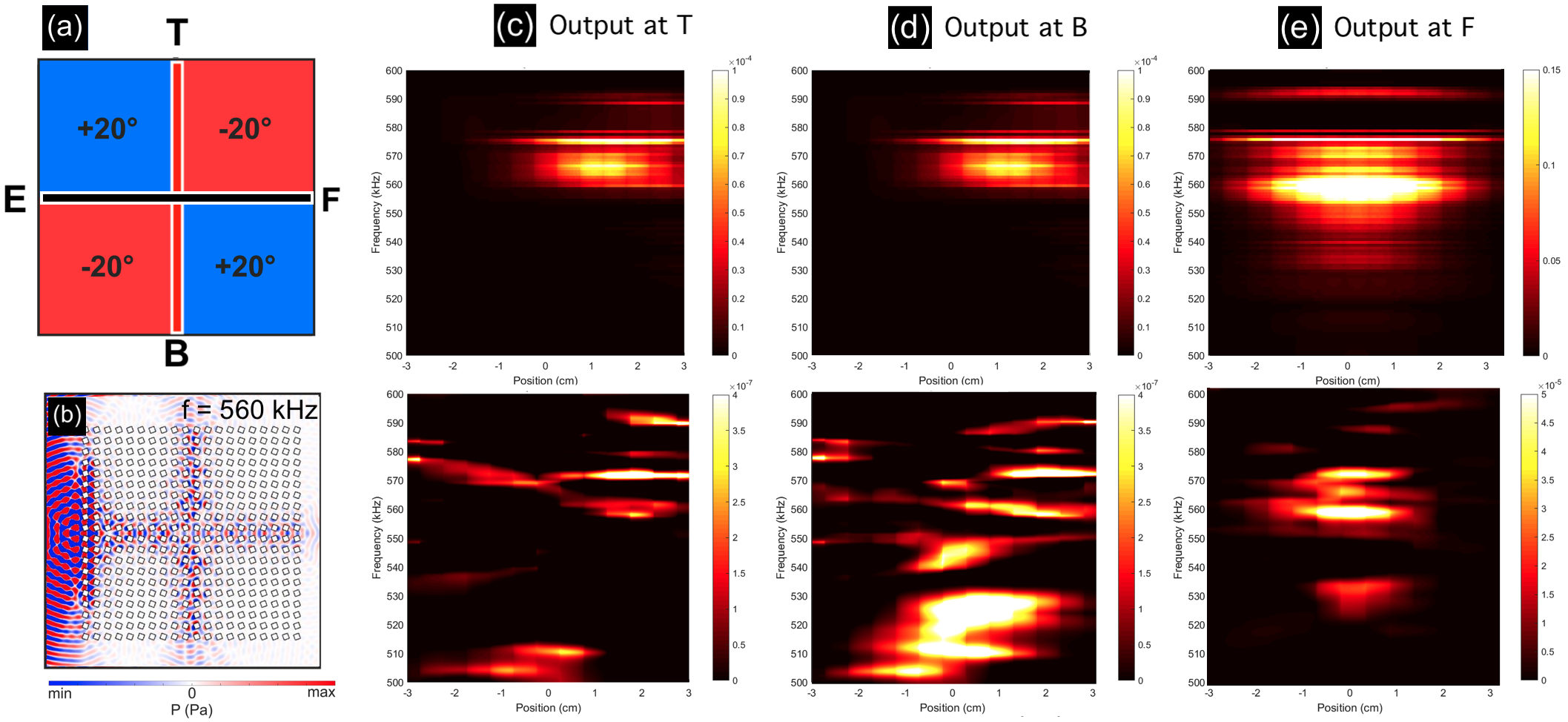} 
	\caption{Three-way splitter for rectangular lattice: (a) shows the four quadrant configuration of our topological domain which resembles Fig. \ref{fig:square_three_way}(a), except now the vertical interfaces (red) differ from the horizontal interfaces (black). A scattering simulation, for frequency $560$ kHz, is shown in (b). Panels (c, d, e) show the magnitude of acoustic energy received by the output ports T, B and F. Numerical simulations are shown in the top row, whilst the lower row are the experimental realisations. There exists a simultaneous frequency range $555:580$ kHz in which all three outgoing ZLMs are ignited.}
	\label{fig:rectangle_three_way}
\end{figure*}
 
Direct comparisons between our numerical computations and experimental results, for the square and rectangular cases, are shown in Figs. \ref{fig:square_three_way} and \ref{fig:rectangle_three_way}, respectively. Importantly for both geometrical cases, there exists a simultaneous frequency range ($550:580$ kHz for the computations and  $555:585$ kHz for the experiments), in which all three output ports receive a transmission. This frequency range lies entirely within the band gap ranges of Figs. \ref{fig:dispersion_curves} and \ref{fig:band_gap}, and hence the excitations in Figs. \ref{fig:square_three_way} and \ref{fig:rectangle_three_way} can be attributed to outgoing ZLMs propagating along the three disparate domain walls. The experimental results closely resemble the numerical results, however there are differences due to possible mismatch in the material constants used for steel and small alignment errors induced by the slender rods not remaining perfectly parallel along the vertical invariant axis. Furthermore, the numerical simulations are two-dimensional and assume invariance along the third dimension, whereas the experimental acoustic beam suffers some natural diffraction spreading. Various airborne acoustic valley-Hall experiments have passing similarity, but require artificial confinement of the field by 
walls, above and below the sample, thereby constructing a waveguide
 \cite{lu_observation_2016}; the approach here requires no such confinement as it is genuinely performed in the bulk and is, hence, more relevant for practical applications.

Despite the spacing between the steel rods for the rectangular case (Fig. \ref{fig:rectangle_three_way}) being a small perturbation away from the square case (difference of $0.02$cm for $L_1$), there are a several marked differences in their transmission properties. The asymmetry of the interfaces results in the vertical leads, which output at T and B in Fig. \ref{fig:rectangle_three_way}(a), being different from the horizontal leads. This difference is evident when comparing the outputs at T and B in Figs. \ref{fig:square_three_way}(c, d) with those in Figs. \ref{fig:rectangle_three_way}(c, d). There is greater excitation in the frequency range $555:580$ kHz which can be ascribed to the shifting of the edge state curves in Fig. \ref{fig:asymmetry_edge_states}. The output at F in Fig. \ref{fig:rectangle_three_way}(e) resembles that in Fig. \ref{fig:square_three_way}(e); this is expected as the outgoing leads upon which the ZLM is hosted is identical in both of these cases. 

These experimental results differ from all of the prior research \cite{cha_experimental_2018, he_acoustic_2016, he_two-dimensional_2019, khanikaev_two-dimensional_2017, nanthakumar_inverse_2019, ozawa_topological_2019, qiao_current_2014, schomerus_helical_2010, shen_valley-projected_2019, xia_topologically_2019, yan_-chip_2018, ye_observation_2017, cheng_robust_2016, wu_direct_2017, xia_topological_2017, zhang_manipulation_2018, qiao_electronic_2011} that have only demonstrated the two-way partitioning of topological modes. Our design is wholly contingent upon the geometrical properties of the square and rectangular lattices. The absence of the diagonal mirror symmetries, $\sigma_d(x_1 \pm x_2)$, for the rectangular model indicate that this may be the model of choice, rather than the square, as it leads to lower modal excitations away from the edge states.


\begin{figure}[htb!]
	\centering
	\includegraphics[width=8.95cm]{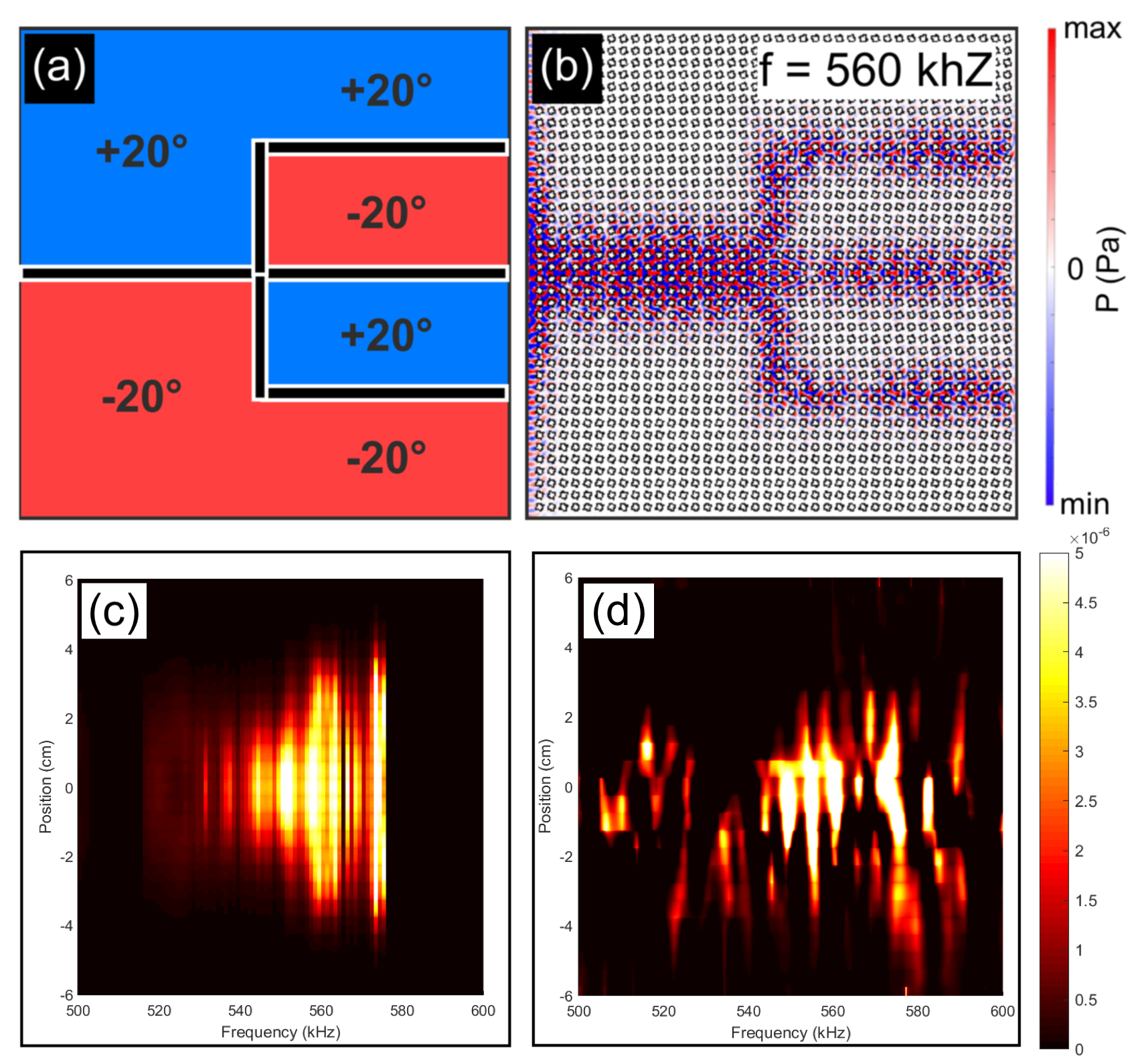} 
	\caption{Trident-like topological network: (a) shows the configuration of the topological domain. (b) shows an exemplar displacement pattern for the frequency $560$ kHz; this network differs from Fig. \ref{fig:square_three_way} due to the alignment of the three outgoing ZLMs that transmit to the same, rather than different, interface. Numerical simulations (c) and experimental realisations (d) show that the frequency range of validity for this exotic effect is $555:580$ kHz.}
	\label{fig:trident}
\end{figure}

\subsection{More advanced topological acoustic circuits}
\label{sec:trident}

By leveraging the results in the earlier section, we design a more complicated topological network, arranged to produce three output ports, that are now aligned along the same edge, Fig. \ref{fig:trident}(a); this design is applied to the square lattice but can be easily extended to the rectangular case. The topological domain, under consideration here, is effectively a composition of the three-way splitter configuration, Fig. \ref{fig:square_three_way}(a), alongside two $\pi/2$ wave steerers. Hexagonal systems cannot produce topological modes that traverse around a $\pi/2$ bend, as this would require a ZLM hosted along a zigzag termination to couple with a mode belonging to an armchair termination. The latter mode, is not topological as it involves the superposition of the $KK'$ valleys and hence the resultant edge states lack a distinguishable pseudospin \cite{ren_topological_2016}; this produces gapped edge states that are, in turn, less robust than their gapless counterparts. The square (and rectangular) lattices do not have this inherent issue and hence our domain in Fig. \ref{fig:trident} is topological \cite{makwana_tunable_2019}. The scattering simulation in Fig. \ref{fig:trident}(b) elucidates the trident-like displacement of the acoustic modes. The numerically computed transmission exists over a range of frequencies, $555:580$ kHz, as seen in Fig. \ref{fig:trident}(c); this range is approximately mirrored by the experimental realisation in Fig. \ref{fig:trident}(d). The three distinct outputs are not clearly discernible in Fig. \ref{fig:trident}(c, d) as the separation between the output ports and the width of the transducer are comparable. Specifically, the transducer is $2.56$cm wide and the output ports are separated by $3$cm; this results in significant convolution as the transducer essentially measures the output from two ports rather than one.  Despite this, there is a higher intensity visible at position $0$ cm, when compared with the magnitude at $\pm 3$ cm, this indicates that there is, most likely, a separation between the three (outgoing) parallel energy channels. We believe this might have potential applications in multiplexing/demultiplexing of acoustic signals, in a way similar to what was proposed for light in photonic crystals with defect lines \cite{centeno_multiplexer_1999}.

\section{Conclusion}
\label{sec:conclusion}

Herein we have shown in detail how to design square and rectangular acoustic topological networks containing energy-splitters that partition energy in more than two directions. The main concepts used to design these systems have been laid out systematically in Sec. \ref{sec:geometry} and the differences resulting from the use of the, not readily studied, rectangular model are highlighted. Our theoretical predictions are followed by experimentally observations in Sec. \ref{sec:experiments} where we experimentally demonstrated the existence of complex topological valley-Hall transport for underwater acoustic waves within non-hexagonal structures. These demonstrations open up a useful way for design in energy transport: the conventional symmetry constraints associated with hexagonal structures can be relaxed, leading to richer designs of waveguiding networks within phononic systems.

\section*{Acknowledgements}
Nicolas Laforge and Muamer Kadic are grateful for support by the EIPHI Graduate School (Contract No. ANR-17-EURE-0002) and by the French Investissements d’Avenir program, project ISITEBFC (Contract No. ANR-15-IDEX-03). Richard Craster, Mehul Makwana and Richard Wiltshaw thank the UK EPSRC for their support through grants EP/L024926/1, EP/T002654/1 and EP/L016230/1.


%

\end{document}